# Development of Sleep State Trend (SST), a bedside measure of neonatal sleep state fluctuations based on single EEG channels

Saeed Montazeri Moghadam MSc*[a, b], Päivi Nevalainen MD, PhD[a], Nathan J. Stevenson PhD[c], Sampsa Vanhatalo MD, PhD[a, b]

[a] BABA Center, Department of Clinical Neurophysiology, Children's Hospital, Helsinki University Hospital, Helsinki, Finland.
[b] Department of Physiology, University of Helsinki, Helsinki, Finland.
[c] Brain Modeling Group, QIMR Berghofer Medical Research Institute, Brisbane, QLD, Australia.

* Address for correspondence:
Saeed Montazeri Moghadam
Department of Physiology,
Biomedicum 1, room B129b
University of Helsinki
Haartmaninkatu 8, 00290 Helsinki, Finland.

**Highlights**

1. Monitoring sleep wake cycling is an essential component in neonatal brain monitoring.
2. Detection of quiet sleep epochs is achievable from single EEG channels with deep learning -based methods.
3. Sleep State Trend (SST) can be used to visualize the classifier results in bedside monitors.

**Abstract**

*Objective:* To develop and validate an automated method for bedside monitoring of sleep state fluctuations in neonatal intensive care units.

*Methods:* A deep learning -based algorithm was designed and trained using 53 EEG recordings from a long-term (a)EEG monitoring in 30 near-term neonates. The results were validated using an external dataset from 30 polysomnography recordings. In addition to training and validating a single EEG channel quiet sleep, we constructed Sleep State Trend (SST), a bedside-ready means for visualizing classifier outputs.

*Results:* The accuracy of quiet sleep detection in the training data was 90%, and the accuracy was comparable (85-86 %) in all bipolar derivations available from the 4-electrode recordings. The algorithm generalized well to an external dataset, showing 81% overall accuracy despite different signal derivations. SST allowed an intuitive, clear visualization of the classifier output.

*Conclusions:* Fluctuations in sleep states can be detected at high fidelity from a single EEG channel, and the results can be visualized as a transparent and intuitive trend in the bedside monitors.

*Significance:* The Sleep State Trend (SST) may provide caregivers a real-time view of sleep state fluctuations and its cyclicity.

# 1. INTRODUCTION

Sleep is essential for early brain organization (Allen, 2012, Graven, 2006, Marks et al., 1995, Roffwarg et al., 1966), and a large number of studies have shown the impact of sleep on many levels of neurobehavioral development (Arditi-Babchuk et al., 2009, Ednick et al., 2009, Lam et al., 2003, Mirmiran et al., 2003, Scher et al., 1996, Touchette et al., 2007, van den Hoogen et al., 2017). Clinically, fluctuation of sleep and wake states is seen as a key, holistic indicator of healthy brain function, hence monitoring sleep-wake rhythms has become a common practice in early brain monitoring in the neonatal intensive care units (NICU) (Meder et al., 2022, Thoresen et al., 2010, van den Hoogen et al., 2017). It has also been suggested that sleep states should be monitored at the bedside to help in optimizing care procedures in the interest of minimizing disruption during a particular sleep state, or the sleep-wake cycling (SWC) (Dereymaeker et al., 2017a).

The gold standard of sleep state recognition is behavioral observation (Grigg-Damberger et al., 2007), however physiological recordings can also be used and polysomnography (PSG) recording is commonly taken as the gold standard physiological recording method. It includes electroencephalography (EEG) and polygraphy (respiration, muscle tone, eye movements) signals, implying technical challenges, especially in bedside NICU use. Therefore, methods have been developed to allow an automated sleep assessment from the EEG signal alone (Ansari et al., 2018, Ansari et al., 2020, Ansari et al., 2022, Dereymaeker et al., 2017b, Fraiwan and Alkhodari, 2020, Fraiwan et al., 2011, Ghimatgar et al., 2020, Hsu et al., 2013, Koolen et al., 2017, Pillay et al., 2018, Piryatinska et al., 2009). These methods usually aim at recognizing the two cardinal sleep states of a neonate, active sleep (AS or rapid eye movement sleep, REM) and quiet sleep (QS or non-REM sleep); some works have even aimed at recognizing more sleep states, including separate detection of the wake state (Ansari et al., 2020, Fraiwan and Alkhodari, 2020, Fraiwan et al., 2011, Hsu;, 2013, Pillay et al., 2018). Most of the existing sleep state detectors are trained using feature-based classifiers (Dereymaeker et al., 2017b, Fraiwan et al., 2011, Ghimatgar et al., 2020, Hsu;, 2013, Koolen et al., 2017, Pillay et al., 2018, Piryatinska et al., 2009), and they generally show performance that comes fairly close to PSG detection. However, reasonably high classifier performance has required use of multiple EEG signals and/or temporal smoothing, both of which challenge the utility of such solutions in the typical limited-channel EEG monitoring that prevails in the NICUs worldwide. Some studies using convolutional neural networks (CNNs) have shown improvement in automatic neonatal sleep state classification, even with fewer EEG channels (Ansari et al., 2018, Ansari et al., 2020, Ansari et al., 2022).

Here, we aimed to develop a sleep state classifier that meets the essential needs of NICU implementation, i.e. the classifier should recognize EEG between active and quiet sleep with high enough accuracy using a single channel EEG only; further, we purposed to construct a classifier that would allow an intuitively interpretable visualization, a Sleep State Trend (SST), to be implemented in the bedside EEG monitors. To this end, we employed convolutional neural networks (CNNs) capable of working on any number of EEG channels, and we trained and tested it using both EEG and PSG recordings. In addition, we tested the performance of our novel classifier against a reference classifier described in (Koolen et al., 2017).

## 2. MATERIALS AND METHODS

### 2.1. *Overview*

An overview of the development and performance testing of the sleep state classifier and the SST is shown in Figure 1.

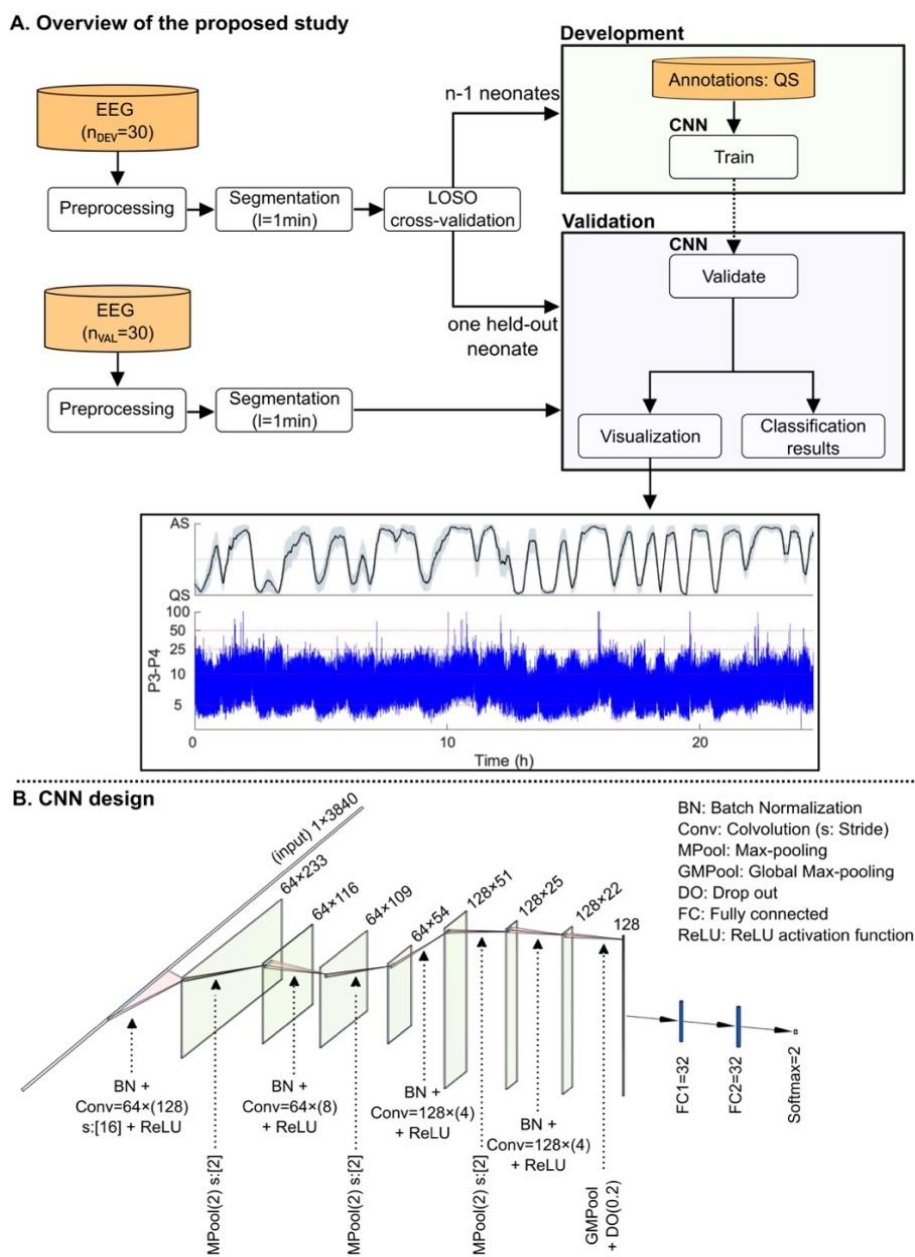

Figure1. **A**. The overview of the sleep state classifier training, its performance assessment, validation, and visualization. **B**. Layer description of the proposed CNN for single-channel sleep state classification with a total of 5,082 parameters. LOSO: Leave-one-subject-out, QS: Quiet sleep, CNN: Convolutional Neural Network.

## 2.2. EEG datasets

**Cross-validation dataset:** We used a cohort of **30** neonates with a total of **943.7** hours of recordings (13 females, gestational age (GA) between 35+1 and 42+1 weeks), all of which exhibited clear sleep-wake cycling in the amplitude-integrated electroencephalography (aEEG) trend. These recordings were collected retrospectively from a clinical database of long-term monitoring (Helsinki University Children's Hospital, Finland) (Nevalainen et al., 2021, Nevalainen et al., 2019). EEGs were collected using a NicoletOne system (Cardinal Healthcare/Natus, USA) at 256 Hz with four scalp electrodes located at F3, F4, P3, and P4, based on the international 10–20 system (Cherian et al., 2009).

The continuous EEG signal was pre-processed with an automated pipeline. First, all the bipolar EEG channels (F3-P3, F4-P4, F3-F4, and P3-P4) were scanned automatically to find and discard high-amplitude values (> ±250 μV) or flat signals (constant zero). In total 118.6 hours of the signal were detected as artifacts and rejected from further analysis. Third, each EEG signal was band-pass filtered at 0.5–30 Hz with a $4^{th}$ order Butterworth filter. Fourth, signals were resampled to 64Hz with an anti-aliasing filter and segmented into 1-minute non-overlapping epochs (discrete signal length of 3840 samples).

This dataset was annotated visually by two experts (SV and PN; hereafter referred to as E1 and E2) for only QS epochs. Since there was only EEG available, the score QS was assigned to epochs with clearly discontinuous, or tracé alternant background pattern. Conversely, the rest was taken as continuous EEG, which may be either AS or awake. This dichotomic scoring may not match perfectly with the vigilance state scoring obtained from a full PSG study, however an approach of this kind has been used in many recent studies, and it is also the underlying assumption in the widely used SWC assessment of the aEEG trends (Thoresen et al., 2010). Notably, the annotations were done at one second accuracy for all the EEG considered to represent QS, i.e. no fixed segment lengths were used. This approach allows physiologically faithful annotation of the QS periods, although it deviates from the clinical tradition of using fixed epoch lengths in the sleep scoring. In total, 175.6 hours of QS (median per neonate: 21 minutes, interquartile range (IQR): 15-27 minutes) were annotated by E1 and 243 hours of QS (median per neonate: 25 minutes, IQR: 18-33 minutes) were annotated by E2. Every 1-minute length EEG epoch was labelled into one of QS or AS scores if all the samples in the epoch belong to the same QS or AS score.

**Independent validation dataset:** The algorithm was validated using an independent dataset that included **30** infants (12 female, GA: 30+3 – 41+1 weeks) with PSG recordings, the gold standard in sleep state classification. These recordings were done using Embla N700 equipment and RemLogic 3.2.0 software (Natus, United States) as per routine clinical protocol, and the dataset was collected retrospectively from the hospital archives using following criteria: the PSG study was performed during newborn period (conceptional age less than 45 weeks), clinical interpretation was normal, and the EEG exhibited tracé alternant as the dominant form of EEG during quiet sleep. Such dataset includes four channels of EEG (C4, O2, A1, Fz) which are not the same recording positions as the training dataset (see above), and hence allow testing how well the findings generalize to EEG in different scalp locations. In addition, the PSG data includes electromyograms (EMGs), electrooculograms (EOGs), electrocardiogram (ECG) and respiratory signals.

The continuous EEG signal was pre-processed with an automated pipeline. First, all the bipolar EEG channels (C4-O2, O2-A1, C4-A1, C4-Fz) were scanned automatically to find and discard high-amplitude values (> ±250 μV) or flat signals (constant zero). Then, each EEG signal was band-pass filtered at 0.5–30 Hz with a 4$^{th}$ order Butterworth filter. Finally, signals were resampled to 64Hz with an anti-aliasing filter and segmented into 1-minute non-overlapping epochs.

The PSG dataset was annotated by clinical sleep specialists as a part of their clinical work, employing the sleep state scoring as per international guidelines established by the American Academy of Sleep Medicine (AASM) (Grigg-Damberger et al., 2007). For the present work, we exported their hypnogram annotations including all the sleep states recognized in our neonatal sleep studies (wake, REM, N1, N3). Here, we tested the algorithm first with a two-class classification where QS was taken to correspond to N3 in the PSG hypnograms, and AS was taken from the other sleep states (wake, REM, N1) grouped together. Notably, such PSG studies include a minimal amount of actual wake, hence the grouping was primarily for REM and N1 states. We also assessed an alternative classification where wake, REM, N1 and N3 were considered separately.

Institutional Research Review Board of the HUS diagnostic center approved the study, including waiver of consent due to the retrospective and observational nature of the study.

## 2.3. Algorithm design

A CNN is a type of artificial neural network (ANN) that is composed of interchanging layers of convolution, nonlinear operator, and pooling. A convolution layer convolves the n-dimensional input tensor and weights matrix. A nonlinear operator (most common ReLU) is then used after each convolution layer to make a nonlinear hidden layer for the CNN. A pooling layer performs a down-sampling process on the output volume of its previous layer to control overfitting. These layers automatically extract features. Then, usually fully connected layers follow those layers to perform the classification. An additional type of layers may be applied depending on the problem, such as batch-normalization to standardize the input to a layer (Ioffe and Szegedy, 2015), drop-out to help prevent overfitting and increase generalization (Srivastava et al., 2014), and softmax to map the output of the last layer to a probability distribution (Bishop, 1995), etc.

The proposed CNN design in this study receives a single-channel 1-minute EEG segment which is initially resampled to 64 Hz as input and after 11 layers of processing gives a vector of probabilities for AS and QS classes as output. Figure1B illustrates the proposed CNN structure which has 5,082 parameters. Prior work has shown that CNN needs a sufficient number of parameters for accurate generalization (Bubeck and Sellke, 2021). The architecture of the network was designed based on some previous studies (Ansari et al., 2018, Ansari et al., 2020, Ansari et al., 2022, Phan et al., 2019a, 2019b, Sors et al., 2018, Zhang and Wu, 2018, Zhang et al., 2020), and trial and error using the development data with a focus on accepting a single-channel EEG as input. Compared to the 1D-CNN solution in prior study, our design has higher number of parameters (Ansari et al., 2018). The implementation was done in Python using Keras with a TensorFlow backend and trained on a Geforce GTX 1070 GPU. The first 8 layers perform feature extraction, and the last 3 layers perform classification and probability calculation.

The optimal CNN classifier was trained with an ADAM solver (beta1 = 0.9, beta2 = 0.99, and learning rate = 0.001) and minibatch size of 64. The parameters were randomly initialized using uniform He initialization (He et al., 2015). Training continued until the validation loss stopped decreasing for a period of at least 35 epochs. Maximum training epochs were set at 500, while most trainings stopped before 100. The learning rate is reduced by the factor of 0.1 once learning plateaus over 20 consecutive epochs.

Post-processing was used for smoothing out spatial and temporal noise in the classifications. The CNN generates an output class probability for each processed 1-minute EEG segment from each channel. This temporal resolution is far higher than what is needed for the NICU brain monitoring purposes where sleep state cycling occurs in the scale of tens of minutes (Curzi-Dascalova et al., 1988, Osredkar et al., 2005, Stevenson et al., 2014) For spatial smoothing, we combined the output probabilities across channels with an averaging function followed by ArgMax to determine the sleep state for each 1min epoch. A very mild temporal smoothing (moving median filter, window length 5 epochs, i.e. 5 minutes) was applied to the combined CNN output in order to taper down incidental noise in the time series.

In the following, we will describe channel-wise results from all four EEG channels separately after the mild temporal smoothing. There was no major difference between the examined four EEG derivations, and each derivation alone gives reasonably accurate results. Channel-wise results are shown to provide a solution for all EEG monitoring contexts, however our post-hoc experiment indicated that spatial smoothing would improve classifier performance (Supplementary Figure 1 and Supplementary Table 1).

### *2.4. Training and performance testing*

Classifier training was done with annotations from both experts; thus, information was incorporated from both the agreements and disagreements, which represent complementary aspects of the experts' annotations, and may improve classifier performance (Moghadam et al., 2021). We excluded epochs with multiple sleep states. The classifier produces one list of two estimated class probabilities (output of the Softmax layer) for every 1-minute non-overlapping EEG epoch from a single channel. For each epoch, the class with higher probability was taken as the prediction.

Classifier performance was initially tested with the training dataset using leave-one-subject-out (LOSO) cross-validation approach, i.e. training with *n-1* neonates and testing on the one "held-out" neonate. This was repeated for all *n* folds. As part of the training process, to prevent overfitting, 10% of the data in the training folds of LOSO was set aside as inner validation (training was stopped if the inner validation loss did not decrease after at least 20 epochs).

For the metrics of classifier performance, we computed confusion matrices (true positive (TP) = correct detection of QS epochs), accuracy, precision, F1 score, Cohen's kappa, and the receiver operator characteristics (ROC, in Supplements) curve. These metrics can be defined with the numbers of true positives/ QS (TP), true negatives/AS (TN), false positives (FP) and false negatives (FN). Accuracy is defined by (TP+TN)/( TP+TN+FP+FN), precision by $TP/(TP + FP)$, F1 score by (2×TP)/(2×TP+FP+FN) and Cohen's kappa coefficient by k = ($P_o$ – $P_e$)/(1 – $P_e$), where $P_o$ represents the observed agreement and $P_e$ denotes the chance agreement.

Clinical utility in individual diagnostics was estimated from the ranges (min-max) of classifier performance measures in individual neonates.

### 2.5. Visualization of the sleep state trend (SST)

Bedside implementation of the developed system requires an intuitive and transparent visualization of the classifier output. As a first step toward such visualization, the clinician should be able to see the classifier output alongside an estimate of its certainty, to be informed of, e.g., ambiguity in the EEG signal for biological or technical reasons. To this end, we propose sleep state trend (SST) by taking a weighted average of the 'probability' outputs from the sleep state classifier for every 1-min EEG epoch. In addition to the SST, an index of classifier uncertainty is added to depict the distribution of classifiers outputs from all channels available in the given recording.

The SST algorithm and its visualization can be used openly via a cloud service by requesting access from the authors.

### 2.6. Comparison to a reference classifier

A recently published feature-based classifier (Koolen et al., 2017) was used as a reference classifier to provide a benchmark with the proposed novel classifier. This allowed comparison of a feature-based approach with our proposed end-to-end deep-learning approach. We also compared our results with this reference classifier after re-training it on our present dataset to the classifier trained earlier with preterm EEG data (67 infants, (Koolen et al., 2017)). Thus, we could assess how well the sleep state detection algorithms could generalize between preterm and term infants. The Koolen's classifier utilizes a set of N=57 computational features combined in a support vector machine (SVM) classifier. The features are extracted from 10-min EEG segments of multichannel EEG, including following categories: (1-2) age-related features, (3-12) frequency-domain features, (13-47) time-domain features, and (48-57) spatial connectivity features. Thus, each multichannel EEG segment contains 228 (57 x 4 channels) features. The SVM classifier was trained and validated using LOSO cross-validation. The hyperparameters of the SVM with radial basis function kernel are set via 6-fold cross validation on the training data.

## 3. Results

For the baseline, interrater agreement between human experts was assessed for the quiet sleep annotations in the full training set. The overall agreement was very high (Cohen Kappa = 0.71, 95%CI: 0.66-0.73) Figure 2A) and represents the upper limit of achievable classifier performance.

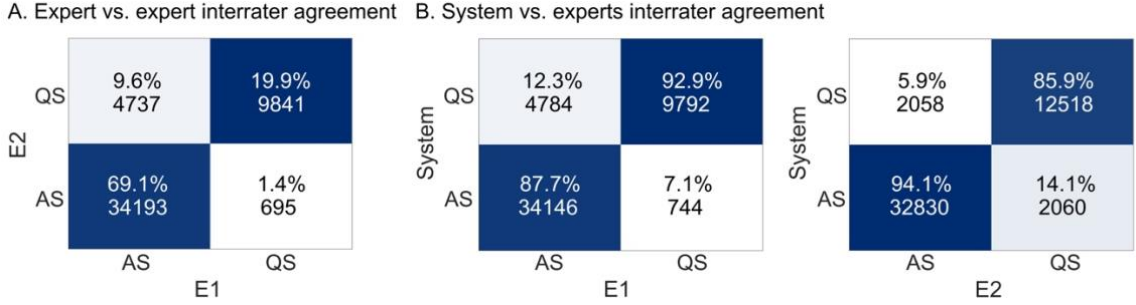

Figure 2. Confusion matrices of agreement of sleep state classification. **A.** between experts. The percentages (and corresponding colours) are with respect to the total number of epochs., **B.** between the proposed classifier on the y-axis and each expert on the x-axis. The percentages (and corresponding colours) denote the recall value of each category. AS: active sleep, QS: quiet sleep.

### *3.1. Algorithm performance on the training dataset*

The performance of the algorithm in detecting quiet sleep was assessed with LOSO cross-validation, using single channel detection (applied to F3-P3, F4-P4, F3-F4, and P3-P4) and the mild temporal smoothing. As shown in Figure 2B, the algorithm performance against each of the experts is overlapping with the interrater agreement between the experts (Figure 2A), and the classifier-expert agreement was very strong as well (Cohen kappa = 0.76, 95%CI:0.73-0.79); Supplementary Table 2). Comparison of EEG derivations showed that the results generalize well across channels (Supplementary Table 1). To further explore the classifier performance vs ambiguity in the EEG, we compared classifier performance levels also against each expert separately, against all epochs, as well as against the epochs with full consensus between experts (disputed epochs (11% of all epochs) are not considered in this case). Finally, the results were compared to the feature-based reference classifier. The results are summarized in Table 1 and Supplementary Table 2. This comparison indicated expectedly (cf. (Airaksinen et al., 2020, Moghadam et al., 2021)) that classification is a bit more accurate when considering the consensus epochs only. Notably, comparison of accuracies in individual neonates showed that detection was a clinically acceptable range (>80%) in all neonates. Performance of the feature-based classifier was generally high as well but the accuracy in individual neonates could fall considerably lower challenging its utility in the clinical workup.

Finally, we estimated how well a straightforward amplitude envelope or its standard deviation computed from the EEG signal could correlate with the SST outputs. As shown in Fig S3, both the amplitude envelope itself and its standard deviation are noisy signals for the purpose of sleep classification, and correspondingly, they only show a week correlation with the SST output.

Table 1. Performance comparison between the reference feature-based algorithm and proposed classifier tested on consensus and all the labels. Since the feature-based classifier is based on multi-channel EEG, number of used epochs for its validation is about one-fourth of number of consensus epochs. PT: feature-based algorithm trained on EEG from preterm (Koolen et al., 2017). T: feature-based algorithm re-trained on EEG from term infants.

|  | # Epochs | Accuracy [%] | F1-score [%] | Precision [%] | Kappa |
|---|---|---|---|---|---|
| Feature-based (PT) | 10005 | 73 (55-86) | 47 (40-56) | 50 (40-57) | 0.2 (0.1-0.3) |
| Feature-based (T) | 10005 | 87 (70-98) | 77 (50-95) | 76 (50-92) | 0.5 (0.3-0.9) |
| Proposed CNN |  |  |  |  |  |
|   Consensus epochs | 44034 | 95 (83-99) | 93 (80-99) | 96 (89-99) | 0.9 (0.6-1) |
|   All epochs | 98932 | 90 (82-95) | 88 (79-92) | 90 (81-95) | 0.8 (0.6-0.8) |

## *3.2. Validation with an external dataset*

To further validate the performance and generalizability of our classifier, we tested it using a PSG dataset, which uses different EEG derivations, a full PSG recording, and has been scored according to the American academy of sleep medicine (AASM) consensus guidelines rather than the EEG background alone (Figure 3). We estimated algorithm's performance in distinguishing N3 (taken to represent QS) from the rest, i.e. wake/REM/N1 (taken to represent active sleep). The overall performance in terms of accuracy was 81% (63-100; the classifier that performed the best in the LOSO cross-validation was used) in N3 versus other sleep states which approaches levels of the inter-rater agreement reported for PSG scoring (Satomaa et al., 2016) (Figure 3A).

We also compared the classifier output against all four PSG-derived sleep states. This analysis (Figure 3B) shows that AS in our detector output most closely represents wake and REM while QS represents N3. There is a confusion between AS and QS for N1; this is expected given the transitory, dynamic nature of N1 state between wake and deeper sleep (Grigg-Damberger et al., 2007).

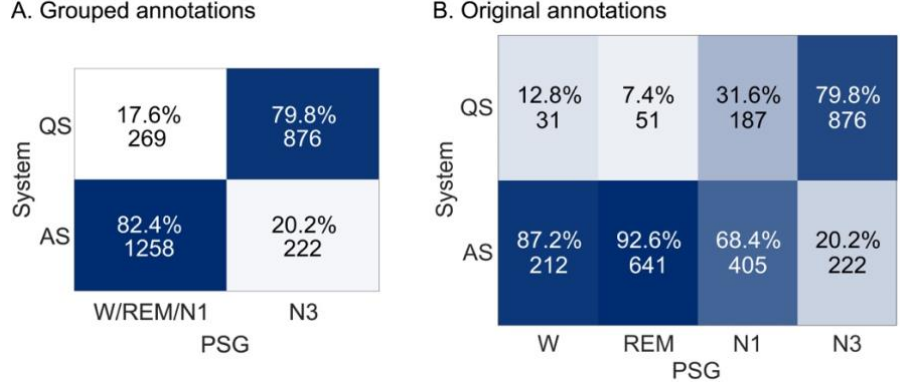

Figure 3. Confusion matrices of agreement of sleep state classification between the proposed classifier on the y-axis and PSG on the x-axis. Numbers are normalized such that each column sums up to one. AS: active sleep, QS: quiet sleep, W: wakefulness, REM: rapid eye movement, N1 & N3: non-REM stages 1 and 3, respectively.

## *3.3. Visualization with sleep state trend (SST)*

In order to be clinically useful, the algorithmic outputs need to be visualized in the bedside monitors. Visualization of sleep state detections is most natural using trends akin to those that are already used in the aEEG displays (Thoresen et al., 2010), seizure detections (Ansari et al., 2019, Stevenson et al., 2019, Stevenson and Vanhatalo, 2018, Tapani et al., 2019), or the vital signs monitors. To this end, we constructed an intuitive SST trend that depicts the weighted average of the probability outputs from the sleep state classifier for every 1-min EEG epoch. A comparable visualization of EEG background (Moghadam et al., 2021) was well received by a representative collection of clinicians. An example of SST is shown in Figure 4 for a 24-hour recording in a neonate from our training dataset. The aEEG trend and the human expert annotations based on the raw EEG are shown for comparison. Sleep state fluctuations are challenging to observe in sections of aEEG trend, especially when it becomes contaminated by the NICU-typical artifacts such as cardiac activity, movements or high-frequency respiration, or when longer time epochs need to be visualized in the same display. The SST is, however, able to depict the sleep state relevant changes with clarity.

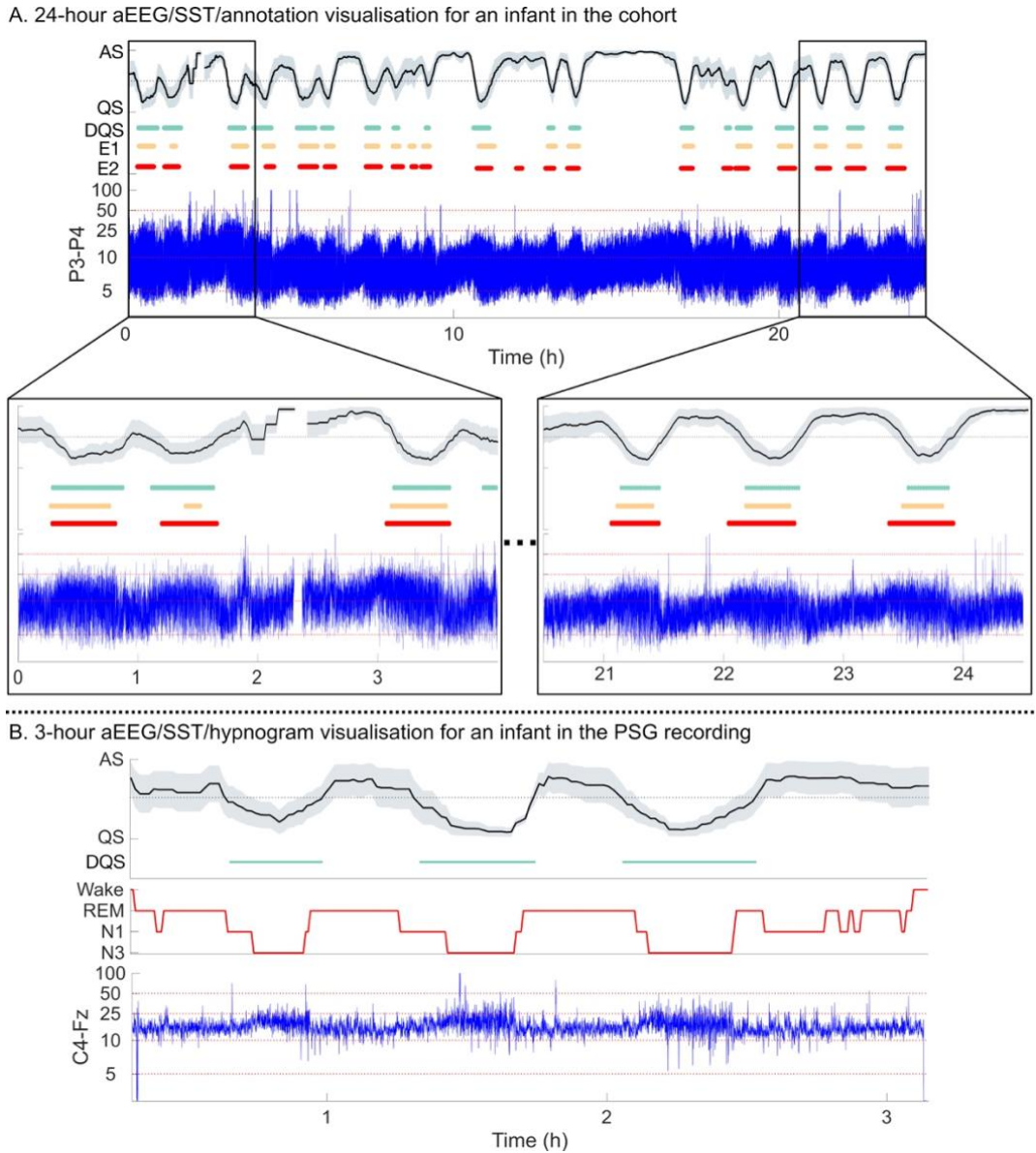

Figure 4. Example SST results. The actual SST (black line) is complemented by depicting the uncertainty of the classifier (gray shadows) at each time point to provide the clinician with a quality index of classification. The uncertainty is quantified by the distribution of the probability outputs of the classifier. Using a fixed threshold (black dotted line in the SST) allows plotting a dichotomic detection of QS states (DQS) which are depicted with green lines under the SST. Expert annotations (E1 and E2) are shown for comparison.

**A**. A full 24-hour signal from the typical P3-P4 derivation. Please note the clear distinction between sleep states in the SST output, with high agreement with the experts' annotations, and a clear change from a poorly organized ("imminent") to a well-organized ("mature") SWC during this time period. This is hard to recognize from the aEEG with 24 hours view, but it is relatively clear in the two 4-hour views of aEEG. The brief discontinuities in the SST output are due to removal of artefactual epochs in the preprocessing stage. **B**. Example of the SST output computed from about 3 hours of PSG recording. The hypnogram depicts the sleep states annotated according to the AASM criteria. Note the overall high agreement between algorithmic detection (SST and DQS) vs hypnogram. AS: active sleep, QS: quiet sleep.

# 4. Discussion

Here we show an accurate detection of QS epochs using a CNN-based classifier on single EEG channel data from newborn infants near term equivalent age. The accuracy of this novel classifier is comparable in EEG channels that are typically used in the long term aEEG monitoring in NICUs. Importantly, the classifier performance generalizes well to an unseen EEG data consisting of other EEG derivations taken from an external validation dataset of PSG recordings; this also provides a gold standard benchmark. Finally, we show that a classifier output of this kind can be visualized in an intuitive manner as a SST trend, which is a continuous display of QS probability and its confidence for quality assessment; importantly, this allows a continuous visual display that can be directly implemented into the bedside (a)EEG monitors. The main novelty of this work is in providing the full pipeline, a complete end-to-end solution from the raw signal to a validated classifier and its implementation into bedside monitors. Moreover, we offer this analytic pipeline openly for any future research use to expedite its take-up into clinical research.

Several prior studies have described automated methods for sleep state detection in the neonatal EEG signals (Ansari et al., 2018, Ansari et al., 2020, Ansari et al., 2022, Dereymaeker et al., 2017b, Fraiwan and Alkhodari, 2020, Fraiwan et al., 2011, Ghimatgar et al., 2020, Hsu;, 2013, Koolen et al., 2017, Pillay et al., 2018, Piryatinska et al., 2009). The earlier solutions were commonly based on computational features extracted from multiple EEG channels and then combined in e.g. SVM-classifiers (Dereymaeker et al., 2017b, Fraiwan et al., 2011, Ghimatgar et al., 2020, Hsu;, 2013, Koolen et al., 2017, Pillay et al., 2018, Piryatinska et al., 2009). Their performance has largely depended on averaging over longer time epochs and combining information from multiple channels in order to show a clinically reasonable accuracy. The need for multiple EEG signals has been a bottleneck in solutions for neonatal EEG monitoring that typically work on one to three EEG channels. More recently, CNN approaches have been developed to provide end-to-end solutions without heuristic feature engineering; methods that have been shown to hold promise in many applications of neonatal EEG classification (Ansari et al., 2020). Previous attempts with CNN algorithms for QS detection have shown that they outperform feature-based methods, which was confirmed in our work as well.

A direct comparison between different classifiers, such as our CNN and feature-based methods, is challenged by their different time resolutions. For instance, our CNN model was trained for 1-min epochs while the Koolen's classifier was initially trained for 2.5-min epochs (Koolen et al., 2017). A shorter epoch length may lead to higher uncertainty in classification, and it is commonly tapered by smoothing in the post-processing phase. For the ultimate implementation in bedside monitors, however, the desired epoch length is defined by the length of sleep wake cycles (SWC); in the neonate, SWC occurs in the scale of tens of minutes (Curzi-Dascalova et al., 1988, Osredkar et al., 2005), hence the epoch lengths between 30s and 10 min used in the classifier constructs are all acceptable to provide measures of cycling (Stevenson et al., 2014) or trends of sleep state fluctuations (Koolen et al., 2017).

Our findings are fully consistent with prior literature showing that QS detection is possible from the neonatal EEG, however we extend previous studies by developing a complete end-to-end solution that takes in single EEG channel data and provides a clinically useful SST visualization. Unlike the prior studies with multichannel data, here we used only single EEG channels, which makes the findings useful as a bedside trend in the (a)EEG monitors. Our findings show an accuracy that compares well with the human interrater agreement, which sets the upper boundary of achievable classifier performance in tasks where the target label is based on subjective, visual interpretation. There are many possible reasons for the improved performance of our novel quiet sleep classifier compared to prior works. For instance, we used a large dataset and annotated the EEG signals in our training data without pre-fixed epochs, which was used in prior studies following the conventional and physiologically imperfect practice in sleep medicine. Larger datasets permit training of larger CNN architectures which increases the potential of the CNN to generalize to unseen data (cf. (Bubeck and Sellke, 2021)). We also incorporated annotations from multiple experts rather than consensus or single annotations as was done previously. Such inclusion of both experts' agreements and disagreements provides a more complex, but accurate, labelling of the data and has been shown to improve classifier performance (Airaksinen et al., 2020, Moghadam et al., 2021). We show here that the present algorithm generalizes well between derivations in the training dataset, as well as to completely different derivations, recording systems and sleep scoring systems used in our independent validation dataset.

In addition to describing the novel QS detector algorithm, we also present an intuitively interpretable visualization of the classifier output, SST, which allows direct implementation in the bedside EEG monitors. SST is not equivalent to the hypnogram that is generated by a human reviewer using international guidelines (Grigg-Damberger et al., 2007), but it allows a simple estimate of sleep cycling between AS and QS which may be clinically useful in many situations (Kidokoro et al., 2012). Unlike prior classifiers trained on healthy neonates, our classifier was trained on neonates that had recovered from serious conditions better reflecting the clinical situation we envision for the SST; tracking evolving SWC from imminent to mature patterns (see Figure 4).

A future clinical implementation of our work needs to consider two limitations: First, the classifier only works for EEG recording from infants at near term age. As shown with our comparison to the reference classifier (Table 1), the sleep detectors may not generalize well between preterm and term age EEG records due to dramatic developmental changes that take place during this period (Bourel-Ponchel et al., 2021, Dereymaeker et al., 2017a, Stevenson et al., 2020, Vanhatalo and Kaila, 2006). Ideally, such an omnipotent classifier could be developed in a hierarchical fashion where the infant's age is first analyzed using the Functional Brain Age algorithm (FBA, (Stevenson et al., 2020)), followed by a sleep state detector that uses FBA as an input. Development of such flexible classifier would be essential to support wide-scale prospective studies on sleep-oriented care and the use of sleep as a functional biomarker across the wide range of conceptional ages present in the NICUs. Second, ambiguity in sleep states may be even more pronounced than in our dataset when monitoring neonates after recovery of cerebral injury, such as birth asphyxia (Thoresen et al., 2010). Evolution of the gradually emerging SWC is considered to be of key interest, however it also presents a conceptual challenge to define sleep stages that emerge from a discontinuous overall background activity. These issues go beyond algorithms, and require characterization by the neonatal community before its faithful detection can be requested from computational algorithms.

A future validation effort that includes prospective data collection, clinically relevant contexts, and a consensus-based definition of conclusions that should be drawn from different means is needed. To make this possible, we present our classifier openly via our Babacloud server (www[dot]babacloud[dot]fi) to anyone interested in its exploitation in clinical trials. Future research is also needed to define the relative importance of detecting quiet sleep vs other vigilance states, and an accurate, specific measure of other sleep states will require corresponding detector development. Detection of any sleep state is sufficient for the purpose of estimating sleep-wake cyclicity (Kidokoro et al., 2012, Stevenson et al., 2014). Since visual review of intermittent QS epochs in the aEEG trend has become the bedside method of choice, our SST trend allows a smooth transition between aEEG and SST interpretations.

When assessing the clinical or scientific utility of a novel method, it is important to consider the potential ambiguities in human annotations, in the data, as well as in the scoring systems; moreover, these should be considered with respect to the aimed implementation. Here, we aimed to develop a classifier that could provide a bedside index of SWC. To this end, it is necessary to have a reliable enough detection of at least one clear sleep state, i.e. QS in our case. Prior literature has shown that QS in the near term infants is mostly recognized as a mixture of tracé alternant and/or high voltage synchronous EEG pattern with an imperfect but good enough accuracy (Andre et al., 2010, Dereymaeker et al., 2017a), and conversely, tracé alternant pattern or its aEEG equivalent has become the hallmark of QS in the clinical work on SWC in the NICU monitoring (Kidokoro et al., 2012, Schwindt et al., 2015, Thoresen et al., 2010). We show that our QS detector works with an accuracy that is equivalent to the level agreement of the human experts visual interpretation in both the training and validation datasets. Hence, the difference between the algorithm's output and an individual human annotations may simply reflect ambiguity in the data itself as much as an error in the algorithm's output. This ambiguity in sleep states becomes even more apparent when considering the spectrum of elusive states that characterize newborn brain recovery from critical illness, such as birth asphyxia (Thoresen et al., 2010). Therefore, the ultimate clinical utility of a novel method like SST should not be evaluated only by a strict comparison with the conventional laboratory methods; rather, one should estimate its perceived added value to the bedside clinician, a work that needs future prospective trials in multicenter settings (Pavel et al., 2020).

**Funding**

This work was a part of a project that has received funding from the European Union's Horizon 2020 Research and Innovation Programme under the Marie Skłodowska-Curie grant agreement (SM: No. 813483), the Finnish Academy (SV: No.313242, No.288220, and No.3104450), Finnish Pediatric Foundation (Lastentautiensäätiö) (SV), Aivosäätiö (SV), Sigrid Juselius Foundation (SV&PN), HUS Children's Hospital (SV), HUS diagnostic center research funds (PN), the National Health and Medical Research Council of Australia (SNJ: APP1144936 and APP2002135).

**Conflict of interest**

Authors declare no conflict of interest.

## Author Contributions

SM and SV conceived the experimental design. PN and SV performed the visual annotation. SM developed the CNN classifier designs, implemented it and analyzed the EEG data. NS developed the original version of the feature-based reference classifier. SM and SV prepared the manuscript and figures. All authors carefully reviewed, commented, and finally approved the article.

# SUPPLEMENTARY MATERIAL

## *Effects of Post-processing*

Here, the overall performance of the proposed classifier on the development dataset before and after post-processing is presented. A comparison between individual bipolar channels versus the combination of channels' output is presented in Table S1. Results showed no noticeable spatial variation, however, combining all the channels yielded a relatively higher performance in all the studied measures. Therefore, the proposed classifier is able to perform at the single-channel level without losing considerable performance.

Table S1. Comparison between individual bipolar channels and combination of the channels.

| Characteristics | F3-P3 | F4-P4 | F3-F4 | P3-P4 | Combined |
|---|---|---|---|---|---|
| Accuracy [%] | 85 (70-93) | 86 (77-91) | 85 (70-91) | 86 (78-93) | 89 (79-94) |
| F1-score [%] | 82 (65-90) | 83 (72-87) | 81 (67-89) | 82 (70-90) | 86 (76-92) |
| Precision [%] | 85 (73-92) | 85 (70-90) | 85 (68-90) | 84 (70-90) | 88 (78-93) |
| Kappa | 0.6 (0.5-0.8) | 0.6 (0.5-0.7) | 0.6 (0.5-0.8) | 0.7 (0.6-0.8) | 0.7 (0.6-0.8) |

Figure S1 shows the improvement in classifier performance after temporal smoothing applied to classifier outputs to remove incidental noise.

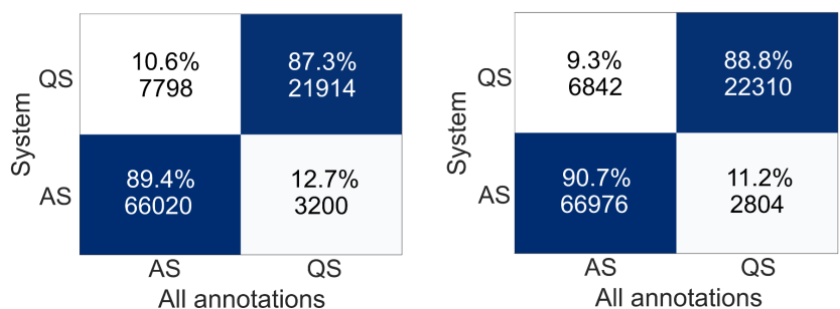

Figure S1. Effect of smoothing on the classifier's performance. **A**. Before post-processing, **B**. After temporal smoothing with a moving median filter (window length = 5 epochs, i.e. 5 minutes). Y-axis: Output of the proposed classifier. X-axis: Combination of experts' annotations. Percentages are normalized such that each column sums up to one.

*Extended performance results on the training dataset*

A version with more details of Table 1 can be found in Table S2. These two tables show the classifier performance compared against each expert separately, against all epochs, as well as against the epochs with full consensus between experts (disputed epochs (11% of all epochs) are not considered in this case). Finally, the results were compared to the Koolen's feature-based classifier.

Table S2. Performance comparison between the feature-based (reference) algorithm and proposed classifier tested on E1 only, E2 only, consensus, and all the labels. PT: feature-based algorithm trained on EEG from preterm (Koolen et al., 2017). T: feature-based algorithm re-trained on EEG from term infants.

|  | # Epochs | Accuracy [%] | F1-score [%] | Precision [%] | Kappa |
|---|---|---|---|---|---|
| Feature-based (PT) | 10005 | 73 (55-86) | 47 (40-56) | 50 (40-57) | 0.2 (0.1-0.3) |
| Feature-based (T) | 10005 | 87 (70-98) | 77 (50-95) | 76 (50-92) | 0.5 (0.3-0.9) |
| Proposed CNN |  |  |  |  |  |
|    E1 annotations | 49466 | 89 (78-95) | 85 (73-92) | 90 (81-95) | 0.7 (0.5-0.8) |
|    E2 annotations | 49466 | 91 (82-96) | 89 (80-95) | 89 (78-95) | 0.8 (0.6-0.9) |
|    Consensus epochs | 44034 | 95 (83-99) | 93 (80-99) | 96 (89-99) | 0.9 (0.6-1) |
|    All epochs | 98932 | 90 (82-95) | 88 (79-92) | 90 (81-95) | 0.8 (0.6-0.8) |

## ROC curves for the proposed CNN sleep state classifier

Further, the receiver operating characteristic (ROC) curve is computed for each held-out neonate during the LOSO cross-validation process and plotted against E1 and E2 annotations separately in Figure S2. This allows to analyze the effectiveness of the classification without defining a fixed threshold. The mean ROC curve over all LOSO tests and the mean area under the curve (AUC) are then obtained and showed in blue solid line and lower right of each panel.

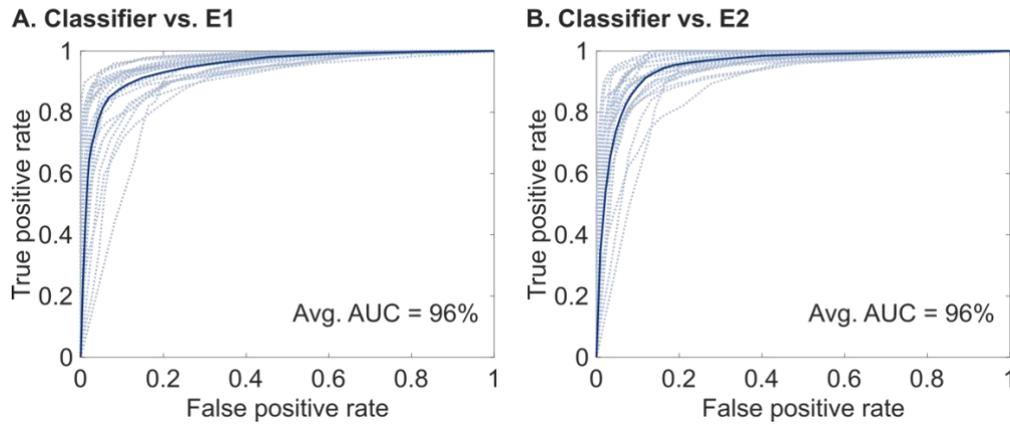

Figure S2. The ROC curves for the proposed CNN sleep state classifier. **A**. Against E1, and **B**. Against E2 annotations. The light blue ROC curves show the performance of the classifier for each held-out neonate during the LOSO cross-validation. The blue solid line represents the mean ROC. The mean area under the ROC curves is equal to 96% for both panels.

*Comparison to a simple envelope detector*

To compare our proposed SST with a plain envelope of the EEG signal, we computed the amplitude envelope from the P3-P4 montage. This was achieved by calculating the Hilbert transform (HT) from EEGs after band-pass filtering between 1 and 30 Hz and then segmenting into 1-min epochs in order to enable direct comparison between the output and the SST. Each EEG segment is given its average (Figure S3. B&C, left) and its standard deviation HT value (Figure S3. B&C, right).

An example visualization of the signal envelope, SST trace, and aEEG trend is shown in Figure S3. A. Despite some co-fluctuation between the envelope and SST time courses, there is a significant "noise" in the envelope which deflates its utility as an index of sleep state. Overall, a very weak positive correlation was observed between SST and envelope values across the full dataset (Figure S3. B; Pearson's $r = 0.14$, $p < 0.01$ and $r = -0.37$, $p < 0.01$). Figure S3. C illustrates the ROC curves and AUCs of the envelope-based methods. According to this figure, an envelope -based sleep state detection is hardly able to exceed the chance level.

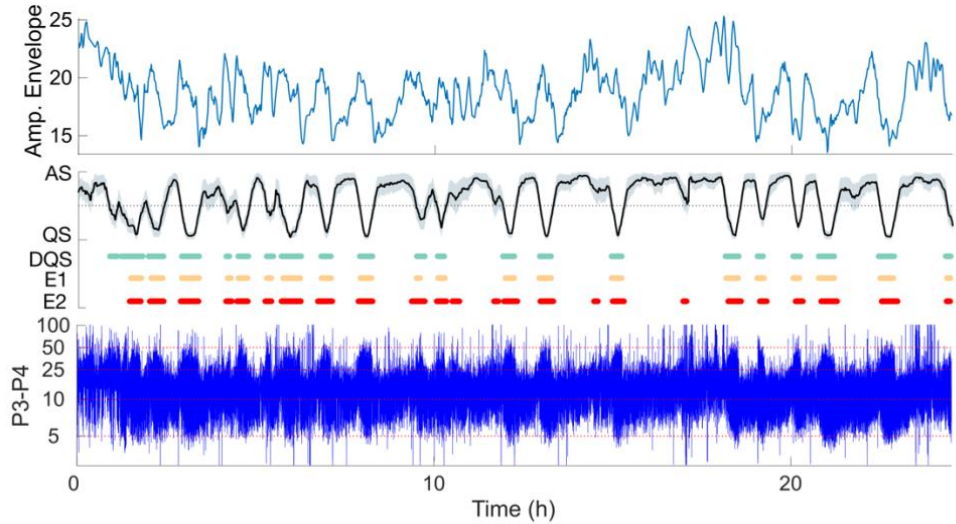
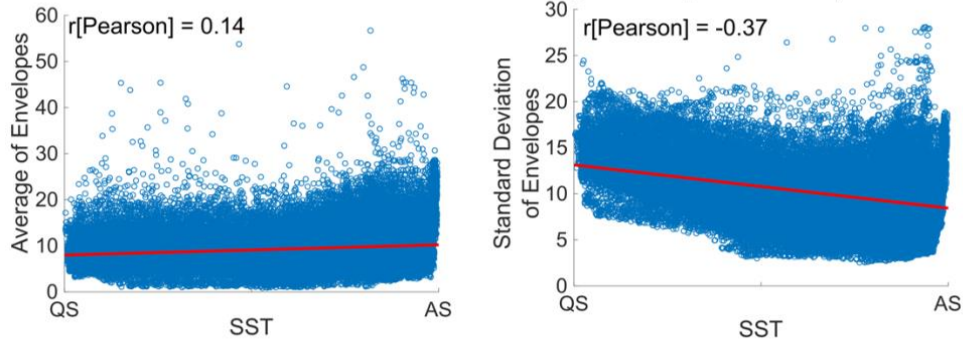
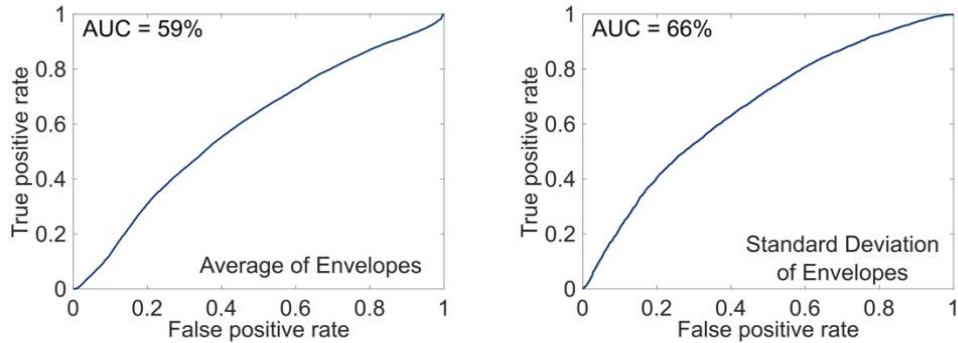

Figure S3. Comparison between SST and a signal envelope in sleep state detection. **A**. An example recording from one infant showing the signal envelope on top, SST trace in the middle, and aEEG trend in bottom. **B**. Scatter plots between SST output and two methods of computing the signal envelope (average; left, and standard deviation; right). A weak correlation (Pearson) is found between envelope-based methods and SST outputs from the corresponding one-minute epochs. **C**. ROC analyses show that the envelope-based sleep state detection performs slightly above chance level.